\documentclass[twocolumn,showpacs,prl]{revtex4}

\ifx\pdfoutput\undefined
\usepackage{graphicx}
\else
\usepackage{graphicx}
\fi

\usepackage[center]{subfigure}
\usepackage{bm}
\usepackage{latexsym}

\def\beq{\begin{equation}}
\def\eeq{\end{equation}}

\begin{document}

\title{Extreme genetic code optimality from a molecular dynamics calculation of amino acid polar requirement}
\author{Thomas Butler and Nigel Goldenfeld}
\affiliation{Department of Physics and Institute for Genomic Biology,
University of Illinois at Urbana Champaign, 1110 West Green Street, Urbana, IL 61801 USA}
\author{Damien Mathew and Zaida Luthey-Schulten}
\affiliation{Center for Biophysics and Computational Biology,
Department of Chemistry and Institute for Genomic
Biology, University of Illinois at Urbana-Champaign, 600 S. Mathews Ave, Urbana, IL 61801 USA}


\date{\today}

\begin{abstract}

A molecular dynamics calculation of the amino acid polar requirement is
presented and used to score the canonical genetic code.  Monte Carlo
simulation shows that this computational polar requirement has been
optimized by the canonical genetic code more than any previously-known
measure.  These results strongly support the idea that the genetic code
evolved from a communal state of life prior to the root of the modern
ribosomal tree of life.

\end{abstract}



\maketitle

The genetic code is one of life's most ancient and universal
features\cite{NIRE63, WOES-book}. It summarizes how RNA transcripts are
translated into amino acids to form proteins, and is shared by all
known cells, across the three domains of life, with only a few very
minor variations\cite{Osawa-codon_capture, Knight-Nat_Rev_Genet}.
Representing a complex series of biochemical steps that comprise all
known cell's translation apparatus, the canonical genetic code is a
mapping from the space of triplets of nucleotide codons to the space
of amino acids. Sequences of codons then correspond to protein
sequences, and ultimately give rise to each organism's phenotype.

Almost immediately after its elucidation, attempts were made to explain
the assignment of codons to amino acids.  It was noticed that amino
acids with related properties were grouped together, which would have
the effect of minimizing translation errors\cite{ZUCK65, SONN65,
WOES1965b}.  In order to determine whether or not this was a genuine
correlation or simply a fluctuation reflecting the limited size of the
codon table, the canonical genetic code was compared to samples of
randomly-generated synthetic codes, starting with early but
inconclusive Monte Carlo work of Alff-Steinberger\cite{ALFF1969}, and
compellingly revisited with larger sample sizes by Haig and
Hurst\cite{HAIG1991}. Depending on the measure used to characterize or
score the sampled codes, high degrees of optimality have been reported.
For example, using an empirical measure of amino acid differences referred 
to below as the ``experimental polar requirement" (EPR) \cite{WOES1966a,WOES1966b}, 
Freeland and Hurst calculated that the genetic code is ``One in a million" 
\cite{FREE1998,HAIG1991,KNIG01}. More recently, it has been shown that 
when coupled to known patterns of codon usage, the canonical code (and the codon useage) 
is simultaneously optimized with respect to point mutations and to the rapid 
termination of peptides that are generated with frame shift errors \cite{ALON2007}.

These results are generally interpreted to imply that the canonical
genetic code had to have undergone a period of evolution, and was not
simply a frozen accident\cite{CRIC1968,Sell06}.  While it was long
assumed that code evolution would be lethal, it has been recently shown
how a genetic code can evolve along with a dynamic refinement of the
precision of translation\cite{Sella_Ardell__no_accident, VETS2006}.
Under vertically-dominated evolutionary dynamics, the optimization of
the code is relatively weak: the system gets trapped in \lq\lq local
minima" and is neither strongly optimized nor converged to a unique
code.  On the other hand, if the evolutionary dynamics is
horizontally-dominated, with genes shared between organisms (as is the
case with contemporary microbes\cite{Ochman}), modularity of structures
such as the translation apparatus and the genome emerges
naturally\cite{Sun07}, and optimization is strong, rapid and convergent
to a universal genetic code, suggesting that early life was communal in
nature and collective in its dynamics\cite{VETS2006}.  Thus, the extent
of optimality of the canonical genetic code is of great interest,
because the greater the level of optimization, the more likely it is
that the genetic code evolved when life was communal in character.

The purpose of this Letter is to set a lower bound on the level of
optimality of the canonical genetic code.  We do this by using
molecular dynamics to construct a measure of code optimality without
any input from experiment, specifically by simulating the equilibrium
behavior of free amino acids in water-pyridine solutions, resembling
those of the original polar requirement experiments, and constructing a
``computational polar requirement" (CPR) by analysis of certain
two-point correlation functions.  We then use Monte Carlo simulation to
determine the level of code optimality, and find that the level is so
high that a new and detailed error analysis is required to ensure
statistically significant assessment of very small probabilities.  We
also explore the dependence of our results on the sensitivity of code
optimality to the scale of code variations.  Our results lend strong
support to evolutionary scenarios for the structure of the genetic
code, with a level of optimization that would only be attainable from
some form of collective dynamics\cite{VETS2006}.  We also report
indications that the dynamics involved the refinement over evolutionary
time of a primitive translation machinery that was ambiguous,
generating a statistical ensemble of related proteins (\lq\lq
statistical proteins")\cite{WOES1965b,woese1973egc} rather than a
unique protein, as is now the case.

\medskip
\noindent {\it Molecular Dynamics of the Polar Requirement:-\/}
The experimental polar requirement is a chromotagraphic measure of amino acid
affinity to a water-pyridine solution that was originally motivated by
a simple stereochemical theory of the origin of the genetic code
\cite{WOES1965a,WOES1965b,WOES1966a,WOES1966b}.  This measure is
related to, and strongly correlated with, several other 
amino acid measures, such as hydrophobicity and Grantham polarity \cite{GRAN74}, but
is not simply related to these other measures. The original polar
requirement experiments used partition chromatography. In the EPR
experiments, water/dimethylpyridine (DMP) ratios ranging from 40-80\%
mole fraction water were used for each amino acid measured.  When the
chromatographic factor, $R_m$ was plotted as a function of mole
fraction water in log-log scale, a linear trend was observed for each
amino acid.  The slope of the corresponding best fit line was taken to
be the amino acid's EPR.

The methods used for obtaining the computational polar requirement
numbers (CPR) have been reported elsewhere \citep{MATH07} and are
summarized here.  The distribution of solute molecules across the
water/DMP interface is related to the equilibrium solvent environment
surrounding the molecules in a binary solution similar to that used in
the experiments.  Trends in the local water density of a solvated amino
acid in water/DMP solutions were found to be linear functions of mole
fraction water.  The slopes of these linear trends were used to obtain
a set of computed CPR values. To quantitatively measure the differences
in local solvent environment, molecular dynamics (MD) calculations were
performed using NAMD2 with an NPT ensemble \citep{NAMD} and the Charmm
27 force field \citep{MACK00,MACK98}.  A pressure of 1.01325 bar and
temperature of 300K were maintained for each simulation.  The systems
consisted of a single amino acid molecule in a box of water and
randomly placed DMP molecules of a determined water/DMP ratio.  For
each amino acid at least four systems, each with a different water/DMP
ratio, were simulated.  Radial distribution functions (RDFs) of water
relative to the amino acid side chains were calculated from the
equilibrated MD trajectories using VMD (Visual Molecular Dynamics)
\citep{VMD}.  The RDFs were calculated by a time average over the
equilibrated portion of a trajectory  \citep{ALLE87}.

\begin{figure}[ht]
\begin{center}
\includegraphics[width=2.5in]{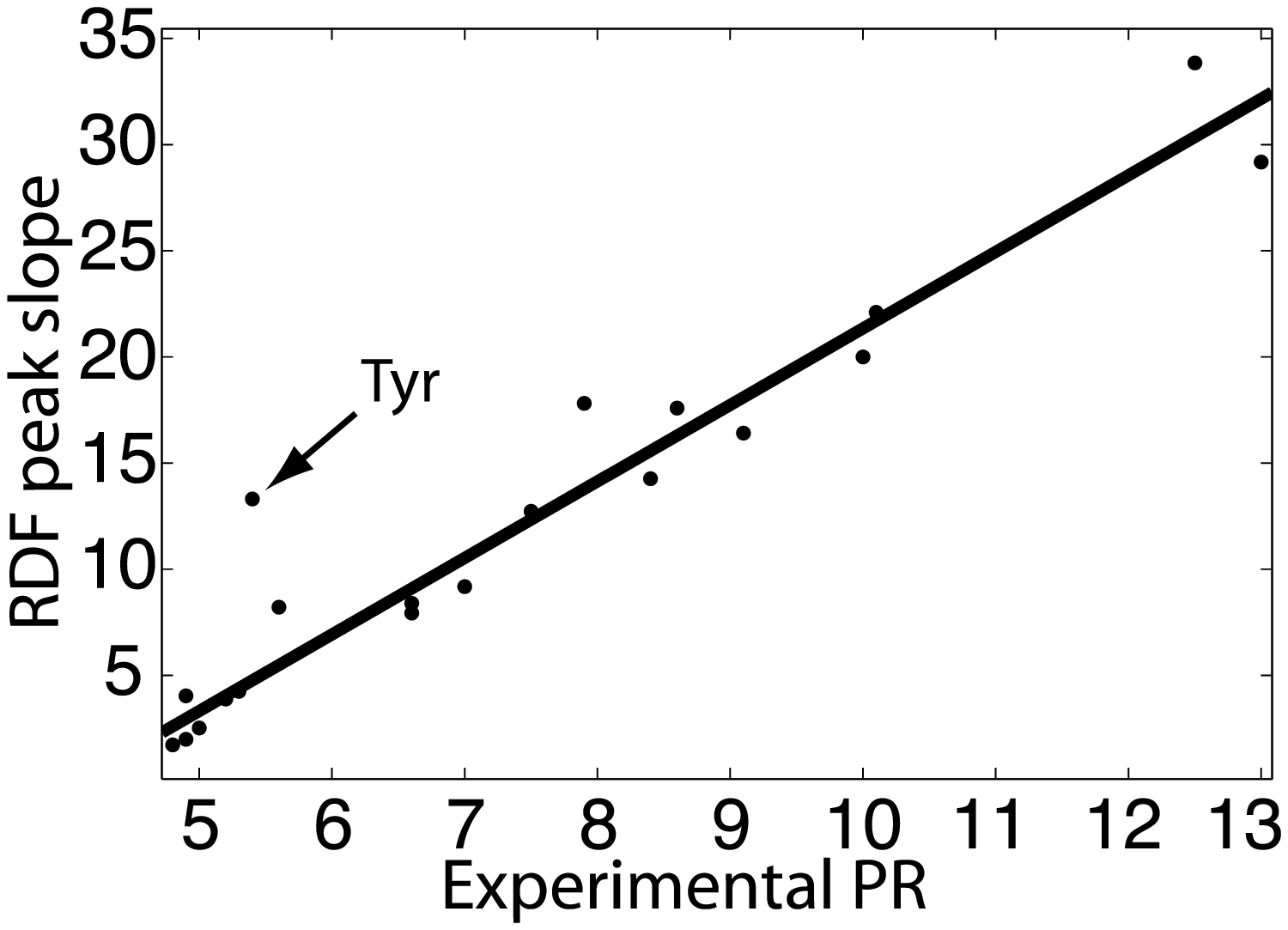}
\caption{Scatter plot showing the relationship between RDF peak slope and experimental polar requirement for all amino acids.  The straight line is a guide to the eye.}
\label{slopefit}
\end{center}
\end{figure}

The most distant atom of the amino acid side chain was used as a
reference atom, and the oxygen or hydrogens (as appropriate) from the
water molecules were used as a selection in calculating the RDFs.
Calculated in this manner, the maximum value of the first peak in an
RDF is related to the relative density of water in the first solvation
layer of the amino acid side chain.  It was found that these maxima
varied linearly with water/DMP ratios for each amino acid, and that the
slopes of the corresponding lines was strongly correlated with the
experimental PR ($R^2 = 0.92$)  (Fig. \ref{slopefit}).
We confirmed that tyrosine's large deviation from the experimental value was not due to a weak signal in the RDF.


\medskip
\noindent {\it Optimality analysis of the canonical genetic code:-\/}
To analyze the CPR, we used the point mutation code analysis algorithms
described in \cite{HAIG1991} and \cite{FREE1998} along with an
analytical realization of bootstrap error analysis to assess the
statistical significance of the results.  The algorithms treat the
genetic code as a mapping $GC^{i}:Codons  \rightarrow  Amino\ Acids$,
where $i$ indexes a particular set of assignments of codons to amino
acids, with $GC^{1}$ as the canonical code. $Codons$ is the set of
codons excepting the termination codons, and $Amino\ Acids$ is the set
of amino acids, i.e. $GC^{1}(UUU)=Phe$.  New versions $GC^{i \neq 1}$
of the mapping are generated by randomly permuting amino acid labels,
leaving termination codons fixed.  This preserves the degeneracy
structure of the genetic code.  The optimality of a given realization
of the genetic code $GC^i$ is assessed by evaluating the sum

\begin{equation}
O_{i}^{-1}=\sum_{\langle c,c' \rangle \neq Ter} {W_{c,c'} \ d^q(GC^{i}(c),GC^{i}(c'))}
\label{1}
\end{equation}
\noindent
where $\langle c,c' \rangle \neq Ter$ denotes a sum over nearest
neighbor codons with the nearest neighbors of a codon defined by its
single point mutations, with all mutations to or from a termination
codon excluded.  The matrix $W_{c,c'}$ weights transition and
transversion biases differently for different positions in the codon,
according to a toy model of typical transversion/transition biases in
real translation.  In our calculations, we used the values from
\cite{FREE1998} as listed in table \ref{table1}. Finally, $d^q(x,y)$ is a
metric on the space of amino acids.  For the polar requirement, the
metric is taken to be $d^q(x,y)=\left| x-y \right| ^q$ over the polar requirement
values corresponding to the given amino acid.


\begin{table}
\caption{\label{table1}The matrix $W_{c,c'}$ of transition/transversion biases taken from \cite{FREE1998}. }

\begin{tabular}{|l|l|l|l|}
 \hline
 &First Base & Second Base & Third base\\
 \hline
Transitions & 1 & 0.5 & 1\\
Transversions & 0.5 & 0.1 & 1 \\
\hline
\end{tabular}

\end{table}

The appropriate quantity to compute is the probability
$P_b=Pr(O>O_{1})$ that a random realization is more optimal than the
canonical code.  To compute $P_b$, we count the number of randomly
generated codes that are more optimal than the canonical code and
divide by the total number of random codes generated.  $P_b$ is
invariant to uniform linear rescaling of the amino acid polar
requirement data, and is smaller for more optimal codes while including
the effects of the large number of codes that can be explored, rather
than the simple linear scale provided by the bare optimality score.

The error in the computed $P_b$ can be estimated using an analytical
realization of bootstrap resampling.  Simulated data sets for bootstrap
are created by randomly sampling optimality scores from the original
data set.  When the samples are drawn from the original set, there are
only two alternatives: a more, or less optimal code can be sampled,
with probability $P_b={N_{O>O_1}}/{N_{total}}$ of drawing a random code
better than the canonical code.  Since the number of better codes in a
sample is the number whose error we wish to estimate, we can regard
drawing a better code as a step to the right with probability $P_b$ in
a one dimensional random walk.  The known formulas for the asymmetric
one dimensional random walk allow us to immediately write down the
bootstrap error estimate in the limit of infinitely many resampled
sets, i.e. the exact bootstrap estimate. For metrics under which the
code is fairly optimal (i.e. $P_b \ll 1$), we obtain the variance in
$P_b$ to be \beq
var[P_b]=var[\frac{N_{O>O_1}}{N_{total}}]=\frac{P_b(1-P_b)}{N_{total}}
\approx \frac{N_{O>O_1}}{N_{total}^2} \label{2} \eeq

To obtain a reasonable estimate of error, or to compare the results of
different metrics on the space of amino acids, the number of more
optimal codes, $N_{O>O_1}$ from the random sample must be sufficiently
large ($\sqrt{N_{O>O_1}} \ll N_{O>O_1}$, or about $N_{O>O_1}=10$ as a
reasonable minimum).

When the computational polar requirement difference squared is used in
the amino acid metric $P_b=(19 \pm 4.36 ) \times 10^{-8}$. In contrast,
with the experimental polar requirement, $P_b=(26.5 \pm 1.63) \times
10^{-7}$, and order of magnitude improvement. To assess the impact of tyrosine (which had the largest variation between the CPR and EPR values) on these results we redid the calculation of $P_b$ for the CPR, but with tyrosine replaced with the value from the EPR. The result is ($P_b=(9.3 \pm 1.0) \times
10^{-7} $).  To test the sensitivity of the results for the CPR, we
varied each element of $W_{c,c'}$ independently by $\pm 0.1 \times
W_{c,c'}$ and repeated the calculation of $P_b$.  This led to results
that were statistically indistinguishable from the results reported
above.  Shorter computations (justified by the faster convergence due
to decreased optimality) for the EPR indicate a similar level of
robustness.  With a $W_{c,c'}$ uniform among nearest neighbors we saw
substantial increases in $P_b$ in agreement with \cite{HAIG1991}.
However, the CPR continued to be superior to the EPR, with the CPR
yielding $P_b=(3.7 \pm .61) \times 10^{-5}$ and the EPR yielding
$P_b=(11.8 \pm 1.1) \times 10^{-5}$.

\begin{figure}[ht]
\begin{center}
\includegraphics[width=3in]{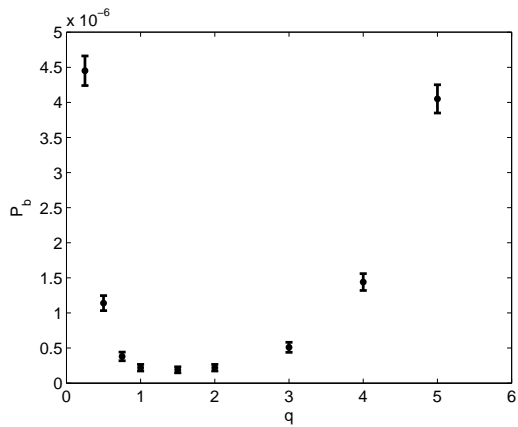}
\caption{$P_b$ as a function of the exponent $q$ in the amino acid metric.}
\label{fig4}
\end{center}
\end{figure}

Varying the value of $q$ in the metric \cite{FREE2000} provides a further probe to
explore the optimization of the genetic code.  Increasing the value of
$q$ is equivalent to emphasizing the role of larger and larger
differences between the amino acid intended, and the one generated by
point mutation.  Thus, if $P_b$ reduces for increased values of $q$,
the code (along with $W_{c,c'}$) evolved to suppress the effects of
rarer, possibly catastrophic errors that may be generated by point
mutations.  This may happen primarily by evolving small elements of
$W_{c,c'}$ where $c \rightarrow c'$ is catastrophic, or vice versa.
Conversely, if $P_b$ reduces for smaller values of $q$, the code
evolved to both mitigate the possibility of these catastrophic errors,
and to minimize the effects of frequent, small errors.  Varying $q$ we
find that the canonical genetic code is most optimal for $q$ between
one and two with significant increases outside this regime in either
direction (Fig. \ref{fig4}). This indicates that the genetic code is
optimized for minimizing errors according to their size with no undue
emphasis to larger or smaller errors.  Given the relative weakness of
the code when emphasizing large errors, evolution must have favored
organisms that discarded or edited fatally flawed proteins over
evolving the code to make them less likely at the cost of reducing its
ability to minimize the more frequent moderate and minor errors.  The
weakness of the canonical code when minor errors are emphasized ($q<1$)
suggests that while the code was still evolving, minor errors were on
the whole less important biologically, as would be expected in 
evolutionary dynamics \cite{VETS2006,Sella_Ardell__no_accident} that utilized ambiguity
tolerance in early proteins \cite{WOES1965b,woese1973egc}.

%
%


\medskip
\noindent {\it Optimality analysis of alternative codes and measures:-\/}
A selection of variant codes were also analyzed using the CPR.
Our findings, displayed in table \ref{table3} were consistent with the
previous findings of Knight in that the alternative codes did not show
marked improvements in optimality over the canonical code
\cite{KNIG01}.  This is consistent with our expectation that
evolutionary pressure to optimize the code with respect to the polar
requirement was eased after the last universal ancestral state.
\begin{table}
\caption{\label{table3} $P_b$ for several naturally occurring variant codes }

\begin{tabular}{|l|l|}
 \hline
Code & $P_b$ \\
 \hline
Canonical & $(19 \pm 4.36 )\times 10^{-8}$\\
Yeast Mitochondrial &  $(11 \pm 3.32) \times 10^{-8}$\\
CDH Nuclear Code & $(21 \pm 4.58) \times 10^{-8}$\\
Ascidian Mitochondrial & $(583 \pm 24.15) \times 10^{-8}$\\
Echinoderm Mitochondrial & $(51 \pm 7.14) \times 10^{-8}$ \\
\hline
\end{tabular}

\end{table}

We also tested Grantham polarity \cite{GRAN74}, which has been argued  
in a survey of genetic code optimality under different amino
acid measures to be the amino acid measure most optimized by the genetic code \cite{KNIG01}.  
The results yield $P_b=(285 \pm 16.88)\times 10^{-8}$, or an order of magnitude
higher than with the CPR metric, leading to the conclusion
that the CPR is the most effective known metric for optimization of the
genetic code.  Previous computations evaluated $P_b$ by generating
$100,000$ random codes \cite{KNIG01}.  Scaling our results to the size
of these original simulations, we see that the EPR and the Grantham
polarity have virtually identical scores.   Scaling the errors for the
CPR and the Grantham polarity to errors assessed from only $100,000$
codes, we get for the CPR, $P_b=(0.19 \pm 0.44) \times 10^{-5}$ and for
the Grantham polarity, $P_b=(2.85 \pm 1.69) \times 10^{-5}$.  These
results are within a standard deviation and a half of each other, and
are therefore not different in a statistically meaningful way.

In conclusion, earlier estimates of code optimality were understated by
a statistically significant amount.  The extent of optimality revealed
here further supports the notion that the genetic code must have
evolved during an early communal state of life\cite{VETS2006}.

We gratefully acknowledge discussions with Carl Woese and thank the
referees for helpful suggestions that improved the manuscript.  This
material is based upon work supported by the National Science
Foundation under Grant No. NSF-EF-0526747.

%
%

\bibliographystyle{apsrev}

\bibliography{optimality}

\end{document}